 \documentclass[final,3p,times,twocolumn]{elsarticle}

\usepackage{amssymb}
\newcommand{\E}{\mathop{\mathrm{ E}}}
\newcommand{\Var}{\mathop{\mathrm{ Var}}}

\journal{\hfill}

\begin{document}

\begin{frontmatter}


 \title{Over Saturation in SiPMs: \\ The Difference Between Signal Charge and Signal Amplitude}
 \author{Max L. Ahnen\corref{cor1}\fnref{label2}}
 \ead{mknoetig@phys.ethz.ch}
 \address{ETH Zurich, Institute for Particle Physics, CH-8093 Zurich, Switzerland}





\begin{abstract}
A recent report on the over saturation in SiPMs is puzzling.
The measurements, using a variety of SiPMs, show an excess in signal far beyond the 
physical limit of the number of SiPM microcells without indication 
of an ultimate saturation. 
In this work I propose a solution to this problem. 
Different measurements and theoretical models of 
avalanche propagation indicate that multiple simultaneous primary
avalanches produce an ever narrower and faster signal. This is because of 
a speed-up of effective avalanche propagation processes.  
It means that SiPMs, operated at their saturation regime, should become faster 
the more light they detect. Therefore, signal extraction methods 
that use the amplitude of the signal should see an over saturation effect. 
Measurements with a commercial
SiPM illuminated with bright 
picosecond pulses in the saturation regime
demonstrate that indeed the rising edge of the SiPM signal gets 
faster as the light pulses get brighter.
A signal extractor based on the amplitude shows a nonlinear behavior 
in comparison to an integrating charge extractor.
This supports the proposed solution 
for the over saturation effect. Furthermore I show that this 
effect can already be seen with a bandwidth of 300MHz, which means that
it should be taken into account for fast sampling experiments. 

\end{abstract}


\end{frontmatter}


\section{Introduction}
\label{sec:sipm}
Silicon Photomiltipliers (SiPMs)
are novel photo detectors designed to detect
weak light signals
and are used for a large variety of applications. Each SiPM consists of a matrix of avalanche diode
cells with a common readout.
The absorbed photons produce hole-electron pairs in the depleted region. Then
the carriers are accelerated 
in the reverse bias electric field and ionize more carriers in an avalanche
breakdown 
process. The avalanche diodes are operated in the Geiger regime. 

Recently, Gruber et al.~\cite{gruber2014} reported on the over saturation in SiPMs.
Their measurements showed an increase in signal beyond the 
physical limit of the number of SiPM microcells without indication 
of an ultimate saturation. They speculated that simultaneous avalanches could
induce a higher charge signal. Here, however, I want to propose a physically more 
plausible model. 

\begin{figure}[ht!]
  \begin{center}
  \includegraphics[width=0.45\textwidth]{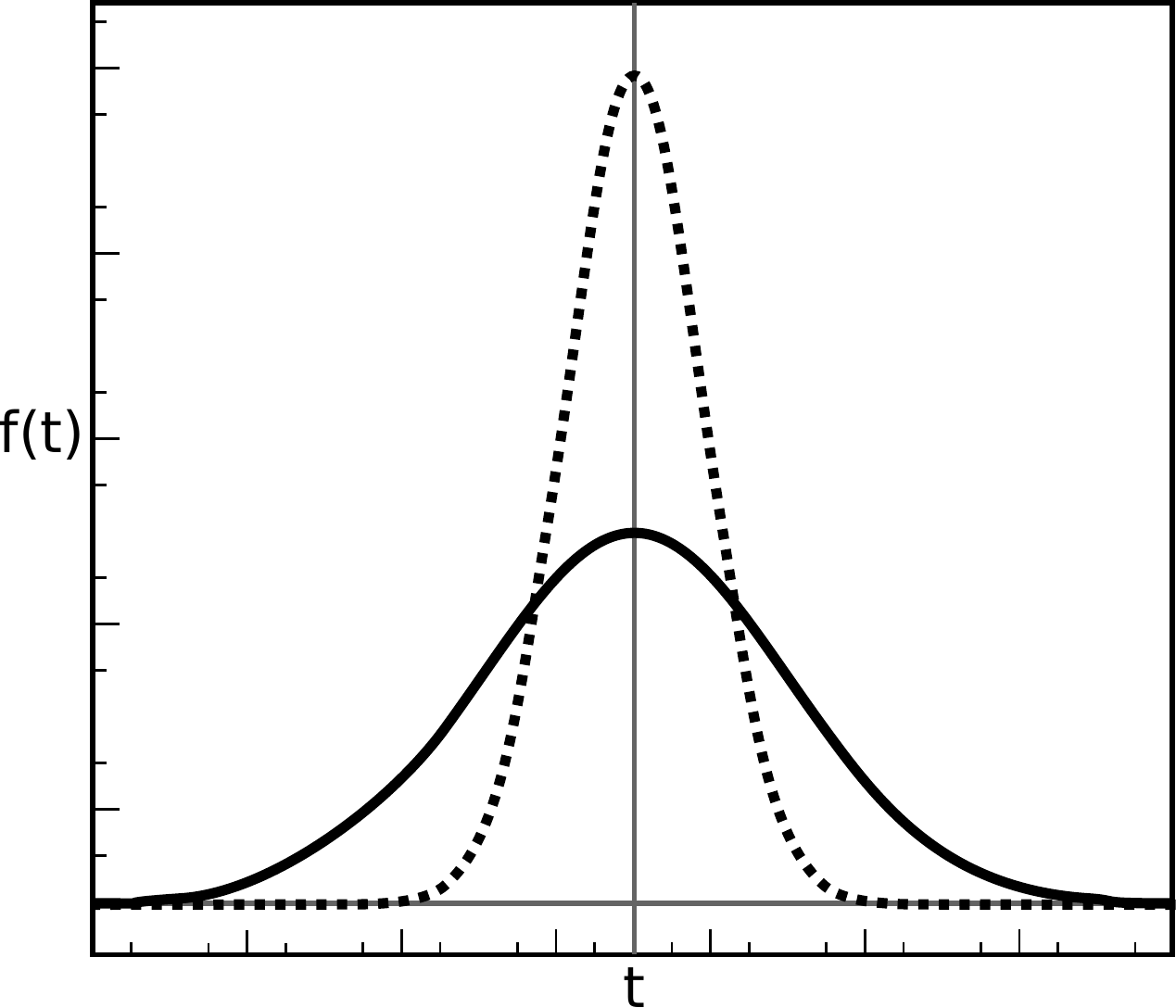}
  \end{center}
  \caption{A schematic of two signals. Both have the 
  same area under the curve (corresponding to the charge), but have 
  different amplitudes, depending on their rise and fall time.}
  \label{fig:dist}
\end{figure}

Theoretical models of 
avalanche propagation~\cite{popova2013} suggest that multiple 
simultaneous primary avalanches produce an 
ever narrower and faster microcell signal. This results from the 
speed-up of effective diffusion processes.
When a photon triggers an avalanche in an SiPM microcell, it grows vertically and laterally
and reaches a maximum lateral size of about $10\mathrm{\mu m}$ after less than 
a nanosecond. This size is usually smaller than the microcell size~\cite{knoetig2014}, 
Therefore the time resolution depends on where in the SiPM 
microcell the avalanche was triggered~\cite{yamamoto2007}.
Popova et al.~\cite{popova2013} indeed 
demonstrate that single microcells produce 
a narrower, higher amplitude pulse when exposed to bright light flashes. Nevertheless 
they observe that the charge within the pulses 
stays the same. Such a behavior can be understood if the 
charge released is constant and the microcell gets faster, see Fig.~\ref{fig:dist}. 

Furthermore it means that the entire SiPM detecting bright, saturating 
sub-nanosecond light flashes should become intrinsically 
faster the more light it detects. This is because every single microcells get faster 
the more simultaneous photons they are exposed to. 

This suggests that charge signal extractors should see normal saturation, while 
amplitude signal extractors should see over saturation --- a signal beyond 
the number of physical microcells. Gruber et al. reported that they indeed used such an amplitude extractor
(the height of the maximum),
to reduce the effect from afterpulsing and late cross-talk. 
They claim to have observed no change in the shape of the signal, yet this 
is what I think is responsible for the over saturation effect they observed. 

\section{Methods and Materials}
\begin{figure}[ht!]
  \begin{center}
  \includegraphics[width=0.45\textwidth]{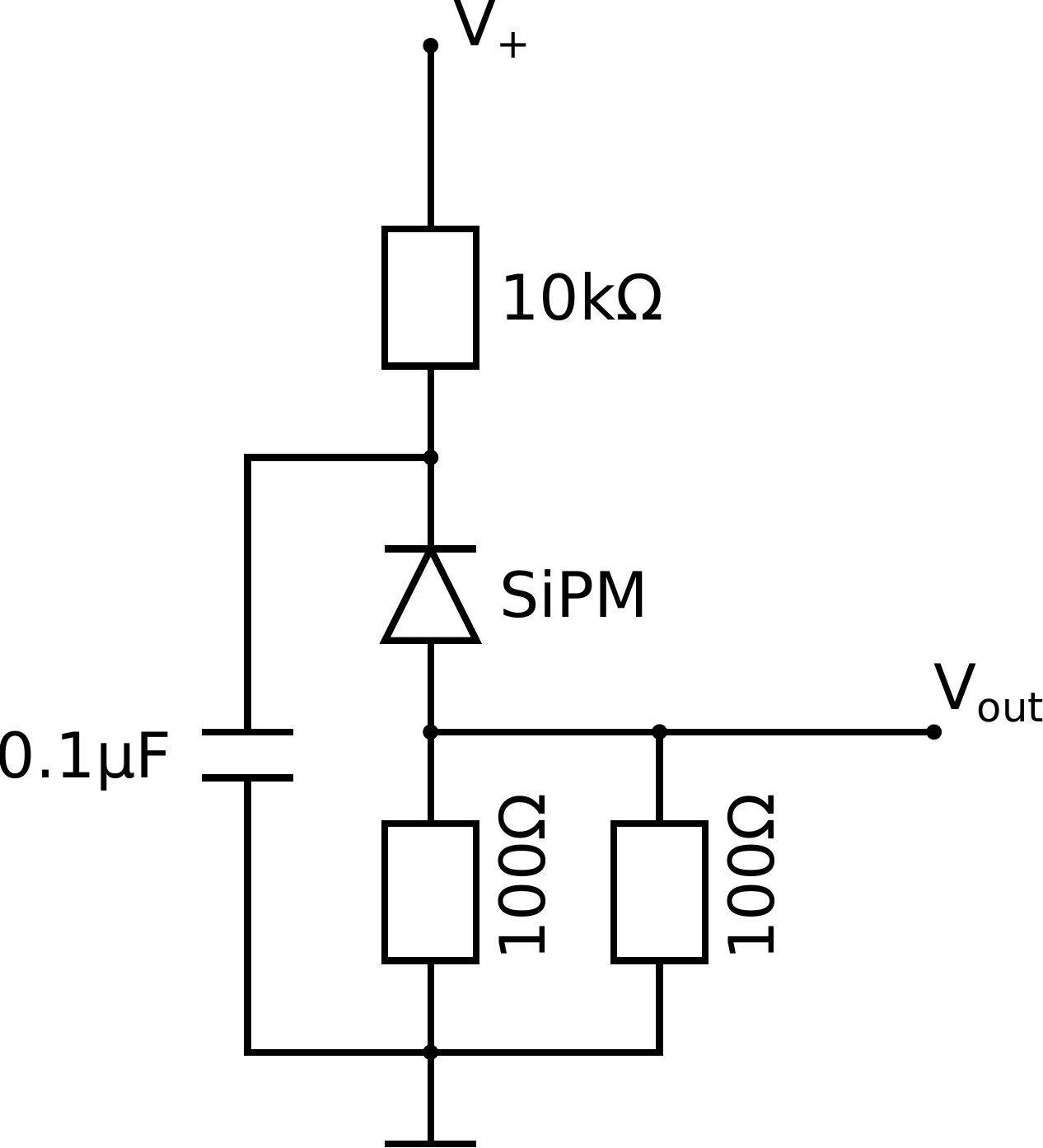}
  \end{center}
  \caption{The schematic of the SiPM board. In order to 
  minimize the distortion of the signal, no preamplifier and no decoupling capacitor 
  are used.}
  \label{fig:schematic}
\end{figure}

To readout the undistorted SiPM signals a  
custom SiPM board, without preamplifier and without decoupling capacitor, is built, 
see Fig.~\ref{fig:schematic}.
Its signals are fed directly into the measurement oscilloscope.
As the measurements are in the saturation regime, the signals from the SiPM 
are large and the intrinsic SiPM amplification is sufficient. 
Also, a preamplifier could distort the measurement. 
The SiPM used is a S10362-33-100C MPPC from Hamamatsu with 900 microcells. 
Another MPPC of the same model is used for verification of the results. 
The bias is set to the data sheet value of $\sim70$V at room temperature. A Picoquant diode 
laser head at 782 nm with a pulse width of few $\sim10\mathrm{ps}$ illuminates the SiPM 
via an optical fiber and an optical diffuser. The laser head intensity
can be continuously adjusted. The fiber, the diffuser, the SiPM and 
the bias board are all placed in a dark box. The fast signals are recorded with a 
LeCroy 2GHz, 10GS/s 204MXi-A sampling oscilloscope. Each waveform is averaged over 
1000 triggers from the pulsed light source which is operated at 1kHz to 
have enough time for the SiPM to recharge. 

As the signals are not amplified, it is unfeasible to calibrate the signals 
via the single photoelectron spectrum. Therefore a calibration via 
the excess noise factor is done. For this I use a Poisson source of 
photons with a mean number of detected photons per pulse $\mu$ and assume a 
geometric chain process cross-talk model with cross-talk probability $p$, 
a random number of detected photons $N$, a random number of detected 
SiPM breakdown cells $X \geq N$, and an excess noise factor $ENF$ defined as the 
loss of signal-to-noise ratio
\begin{equation}
 ENF = \frac{\E^2[N] / \Var[N]}{\E^2[X] / \Var[X]} .
\end{equation}
It follows~\cite{Vinogradov11} that the gain $g$, defined as proportionality constant
between measured signal $Q$ and number of cells in breakdown $X$
\begin{equation}
 Q = g X ,
\end{equation}
can be measured as 
\begin{equation}
 g = \frac{\Var[Q]}{\E[Q]}\frac{1-p}{1+p} .
 \label{eqn:calibration}
\end{equation}
The amplitude extractor gain $g_{amp.}$ was measured at operating voltage using 1000 
low light intensity pulses of $\sim 30$ detected photons/pulse as saturation 
has to be negligible for this measurement. The charge 
extractor gain is estimated by applying a linear model 
to a part of the data where both extractors are proportional to each other (see Sec.~\ref{sec:results}). 
The last step of the calibration is to apply it to 
the mean measured signal $\E[Q']$, in order to determine the 
number
of mean detected photons  $\mu$ in a pulse.
With the simple 
geometric chain cross-talk model, 
the result is
\begin{equation}
 \mu = \frac{\E[Q']}{g} (1-p).
 \label{eqn:calibration2}
\end{equation}
The SiPM under study has $p \sim 0.1$ cross-talk at operating 
voltage and I estimate the accuracy of measuring $\mu$ to $\Delta\mu / \mu \sim10\%$

\section{Results and Discussion}
\label{sec:results}
\begin{figure}[ht!]
  \begin{center}
  \includegraphics[width=0.47\textwidth]{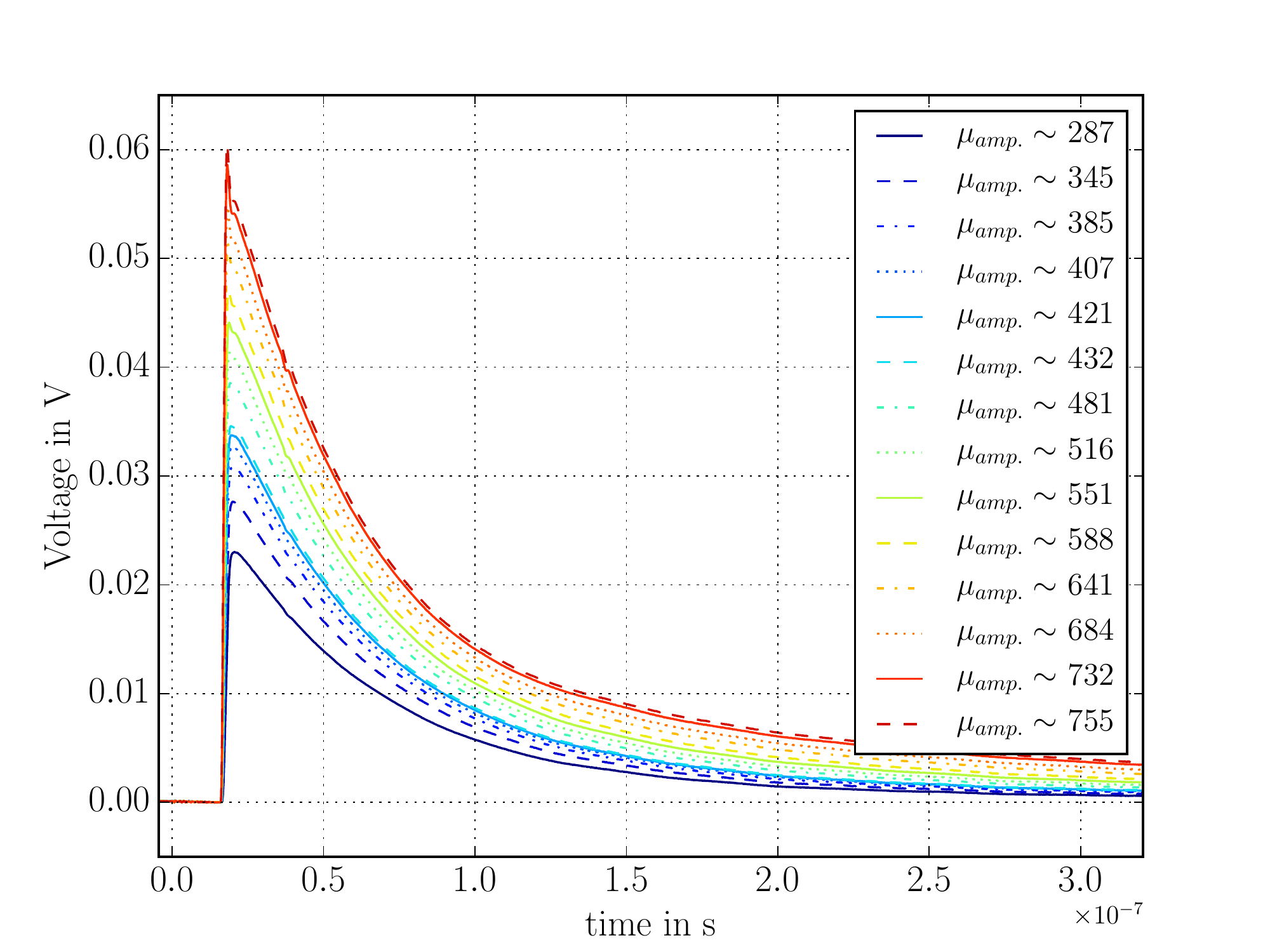} \\
  \includegraphics[width=0.47\textwidth]{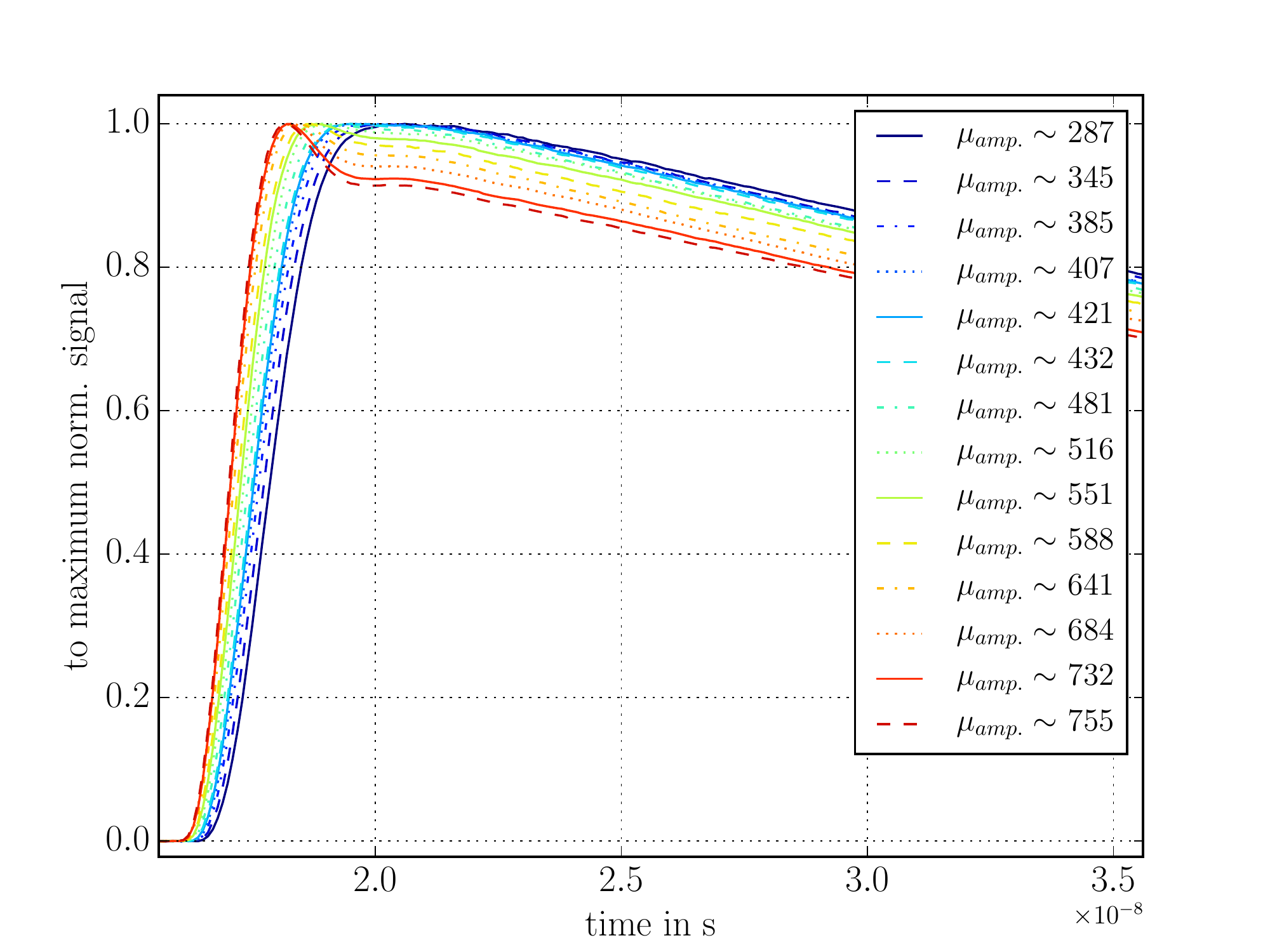} \\
  \includegraphics[width=0.47\textwidth]{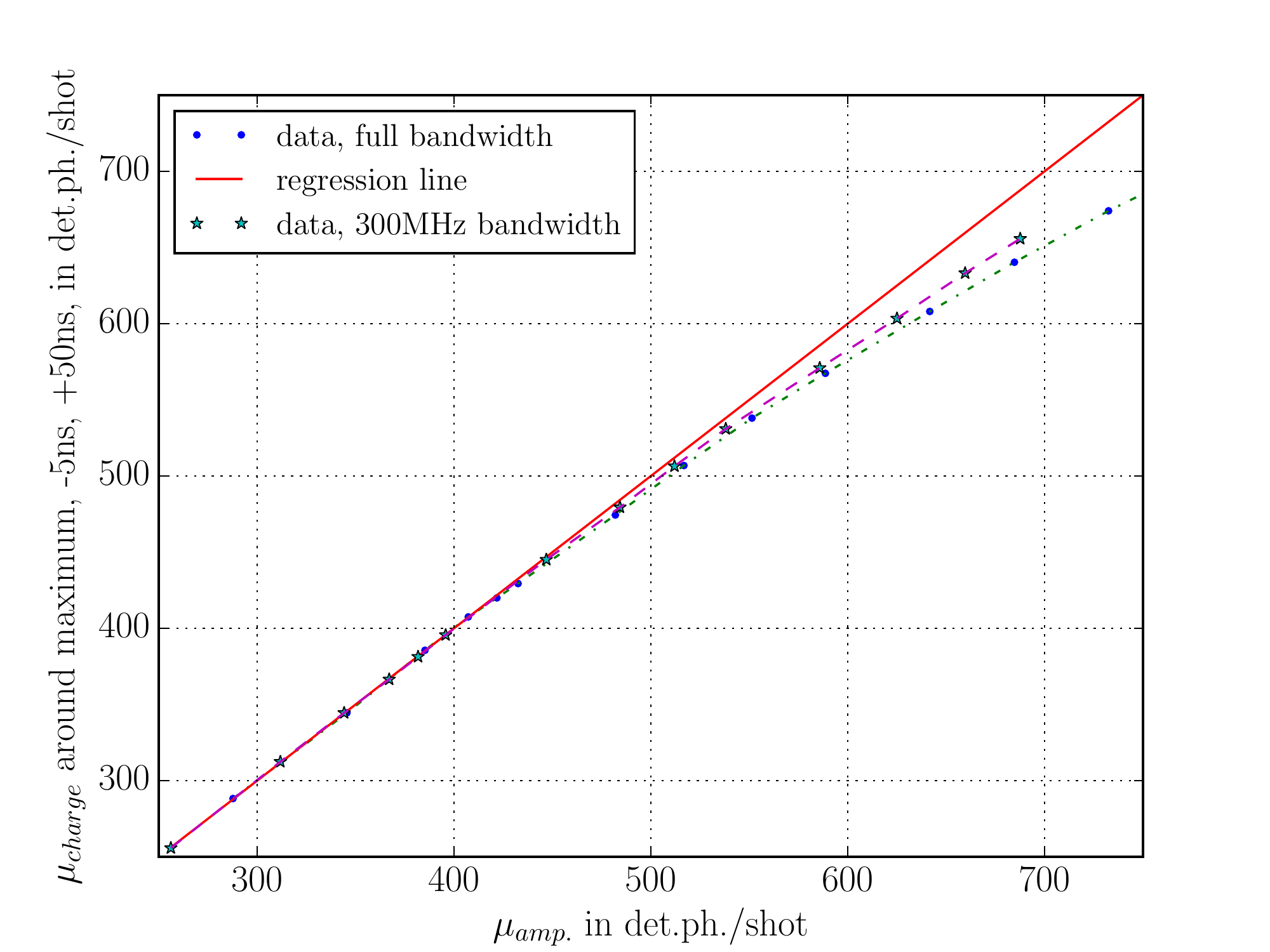}
  \end{center}
  \caption{Top: The oscilloscope signals with full bandwidth.
  Every waveform is averaged and pedestal subtracted.
  The light pulse intensity is increased from $\mu_{amp.} \sim 287$ detected 
  photons per laser shot to $\mu_{amp.} \sim 755$ detected photons per laser shot
  as measured with the amplitude extractor.
  Middle: A zoom into the rising edge of the same oscilloscope signals. 
  They are normalized to their maxima.
  One can see that the rising edge gets steeper.
  Bottom: A comparison of an integrating (charge) signal extractor
  to an amplitude extractor. 
  The amplitude extractor uses the pedestal subtracted maximum, is
  calibrated with Eqn.~\ref{eqn:calibration} \&~\ref{eqn:calibration2}, 
  and detects $\mu_{amp.}$ photons per pulse.
  The charge signal extractor 
  uses an integration window of $-5$ns to $+50$ns 
  around the maximum.
  In order to calculate the charge extractor gain $g_{charge}$, a
  a linear model is fitted to the first four data points on each dataset,
  where both extractors are proportional. This gain is then used to calculate 
  $\mu_{charge}$.}
  \label{fig:measurements}
\end{figure}
Fig.~\ref{fig:measurements} shows the different waveforms as measured with 
the oscilloscope. 
The measurement consists of a series of 14 intensity settings with rising brightness
from about 287 detected photons per laser shot to about 755 detected photons per laser shot.
The waveforms are all pedestal subtracted.
One can see that the rising edges get slightly steeper, 
the brighter the initial light pulse is.
A measurement with a second SiPM of the same type shows the same behavior. 

To quantify this result, Fig.\ref{fig:measurements} also
shows a comparison of
an amplitude extractor and an integrating (charge) extractor.
They are proportional in the lower count regime. 
Both extractors show rising signals, as the physical saturation 
limit of 900 (microcells) is not quite reached. 
Yet, in the saturation regime and assuming the over saturation behavior comes 
from the speeding-up of the avalanche diffusion, one would expect 
that the signal from the amplitude extractor should saturate slower than the integrating 
charge extractor. This non-linear effect is observed. 
The influence 
of afterpulsing on the integrating charge extractor can not 
mimic such a behavior as its effect is to 
increase the charge measurement, opposite to what is observed. 
The figure also shows a measurement conducted with a 
300MHz FIR low pass filter on the oscilloscope. 
This shows that typical fast sampling experiments need to take 
this effect into account (depending on the dynamic range of their 
digitization). 

Diffusion of multiple simultaneous avalanches can further explain how a lower bias leads to more over 
saturation~\cite{gruber2014}, as lower biased cells are slower~\cite{popova2013} and therefore the effect of two 
simultaneous avalanches is stronger. 
It may also explain how SiPMs from the same manufacturer, but different 
microcell size, show different over saturation behavior~\cite{gruber2014}. This is because the 
avalanche has a limited size smaller than a microcell.
Therefore different microcell geometries have different timing behavior. 

\section{Conclusion}
Multiple simultaneous avalanches make the microcells and in turn the whole SiPM
faster, which can explain the previously observed over saturation effect in SiPM.
In this work I report on measurements with a commercial
SiPM from Hamamatsu, which is illuminated with bright 
picosecond pulses in the saturation regime. The measurements indicate 
that the rising edge of the SiPMs gets faster and that 
a fast signal extractor based on the amplitude shows a nonlinear behavior 
in comparison to an integrating charge extractor, supporting the proposed 
explanation. The effect can already be seen with a low pass bandwidth filter of 300MHz.
This means that it should be taken into account for fast-sampling experiments.

\section{Acknowledgments}
I would like to thank R. Mirzoyan for fruitful discussions, 
G. Hughes for comments on the manuscript, and   
L. Ahnen, Biomedical Optics Research Laboratory
- University Hospital Zurich, and M. Kroner, 
Institute of Quantum Electronics - ETH Zurich, for providing the oscilloscope and 
the light source.



\bibliographystyle{elsarticle-num} 
\bibliography{waveshape}



\end{document}